\documentclass[a4paper]{spicawStyle}
\usepackage{graphicx,natbib}

\begin{document}

\title{Spectroscopic Cosmological Surveys in the Far-IR}

\author{L. Spinoglio\inst{1}, M. Magliocchetti\inst{1}, S. Tommasin\inst{1}, A.
M. Di Giorgio\inst{1}, C. Gruppioni\inst{2}, G. De Zotti\inst{3}, 
A. Franceschini\inst{4}, M. Vaccari\inst{4}, K. Isaak\inst{5}, F. Pozzi\inst{6}
\and M.A. Malkan\inst{7}} 

\institute{
Istituto di Fisica dello Spazio interplanetario - INAF, Via Fosso del Cavaliere 100, 00133 Roma, Italy
\and 
Osservatorio Astronomico di Bologna - INAF, Via Ranzani 1, 40127, Bologna, Italy
\and
Osservatorio Astronomico di Padova - INAF, Vicolo dell'Osservatorio 5, 35122 Padova, Italy
\and
Dipartimento di Astronomia - Universit\'a di Padova, Vicolo dell'Osservatorio 5, 35122 Padova, Italy
\and
University of Cardiff - School of Physics and Astronomy, 5 The Parade, Cardiff CF24 3YB, United Kingdom
\and
Dipartimento di Astronomia - Universit\'a di Bologna - INAF, Via Ranzani 1, 40127, Bologna, Italy
\and
Astronomy Division, University of California, Los Angeles, CA 90095-1547, USA}

\maketitle 

\begin{abstract}

We show the feasibility of spectroscopic cosmological surveys with the
SAFARI instrument onboard of SPICA. The work is done through simulations
that make use of both empirical methods, i.e. the use of observed luminosity
functions and theoretical models for galaxy formation and evolution. The
relations assumed between the line emission 
to trace AGN and star formation activity have been
derived from the observations of local samples of galaxies.
The results converge to indicate the use of blind spectroscopy with the
SAFARI FTS at various resolutions to study galaxy evolution from the local
to the distant (z$\sim$3) Universe.

Specifically, two different and independent galaxy evolution models predict about 
7-10 sources to be spectroscopically detected in more than one line in a 2'$\times$ 2'
SAFARI field of view, down to the expected flux 
limits of SAFARI, with about 20\% of sources to be detected at z$>$2.
SPICA-SAFARI will be therefore excellent at detecting high-z sources and 
at assessing in a direct way their nature (e.g whether mainly AGN or Star 
Formation powered) thanks to blind spectroscopy.

\end{abstract}

\section{Introduction}
  
In the last years our perspective of galaxy evolution has greatly changed thanks to 
two main findings: A) the strong correlation 
observed in the local Universe between the mass of the black hole at the 
centre of a galaxy and the velocity dispersion of 
the stellar component of the galactic bulge (the so-called {\it Magorrian
 relation}), and B) the evidence that most, if not all, 
galaxies during their evolution pass through a FIR/submillimetre bright phase.

A) The Magorrian relation \citep{Mag98, Fer00} 
implies that the processes of black hole growth - through mass accretion - and bulge formation - 
through star formation - are intimately linked. 
While large elliptical galaxies with 
old stellar bulges are known to follow the {\it Magorrian relation},
we are not able to explain this tight correlation for active galaxies
in the local Universe, as the two processes appear 
- at least in local Seyfert galaxies \citep{t09} 
to be almost independent of each other. 
The {\it Magorrian relation} therefore has to find its origins at earlier epochs.

The study of galaxy evolution necessarily implies investigation of the full cosmic history of energy 
generation by stars (star formation and stellar evolution) 
and black holes (accretion), as well as of the energy loss processes such as the feedback from AGNs, 
because all these are responsible for the build up of the baryonic mass in the Universe 
and must ultimately 
be consistent and set up the observed local relation between luminosity and mass in galaxies.
To understand the Magorrian relation we need to make the cosmic connections between stars in a 
galaxy and its massive black hole. The global accretion power, measured at X-rays \citep{Has05} 
and the star formation power, measured by H$\alpha$ and rest-frame UV observations \citep{Shi09}
were $\sim$20 times higher at z=1-1.5 than today. On a cosmic scale, the evolution of supermassive 
black holes (SMBHs) appears related to the evolution of the star-formation rate (SFR),  strongly 
suggesting the presence of co-evolution \citep{Mar04,Mer04,Sha09}.
As suggested by various authors \citep{Hec04, Gra04}, the growth of bulges through 
SF may be directly linked to the growth of black holes through accretion. 
Quasars have been advocated as a source of negative feedback that would quench star formation, 
however no clear evidence for this ''negative feedback" has 
yet been found, while instead star formation is often, although not always, 
concomitant with AGN/QSO activity over rather
long duty cycles (100 Myr - few 100 Myr). It has also been realized that the optically 
bright phase of quasars covers not more
than one tenth of their host lifetime, as observations generally constrain quasar lifetimes to the range 
$\sim$ 10$^7$ - 10$^8$ yr. 
Optical studies of local galaxies show that most, if not all, large galaxy spheroids host massive relic black-holes 
\citep{Ric98} which, in turn, suggests that a mass-accreting AGN phase is one through which all galaxies pass.

B) To account for the total energy generated by stars and that by black hole accretion, 
one first has to determine how much of the observed luminosity 
is partly or heavily extinguished (reddened or obscured). We know from both galactic and extragalactic 
studies that the processes of star formation 
and early stellar evolution is usually deeply enshrouded in dust. Active galactic nuclei (AGNs) are often 
optically obscured, both locally e.g. (e.g. \citet{Gou09}) and at high redshifts \citep{HC09}. 
Thus, galaxies pass a significant fraction of their lifetime deeply obscured by dust.

Furthermore, recent observational studies assesses that  during their evolution most (if not all) galaxies pass through a 
FIR/\-submillimetre bright phase, where star formation is probably the dominant energy 
production mechanism, but where also supermassive black-hole accretion may contribute significantly. 
The luminous and ultra-luminous infrared galaxies, 
discovered in the 80's by the all-sky survey of IRAS, are
rare in the local Universe, but they become much more common at high-redshift, as shown by the detection of the 
Submillimeter Galaxies (SMG). Their 
integrated luminosity accounts for a significant fraction of the cosmic infrared background (CIRB) 
\citep{Dol06}, the total energy of which exceeds that of the optical background 
and implies that more than half the star-formation activity in the universe is hidden by dust extinction. 

\label{}
 \begin{figure}[ht]
  \begin{center}
    \includegraphics[width=\columnwidth]{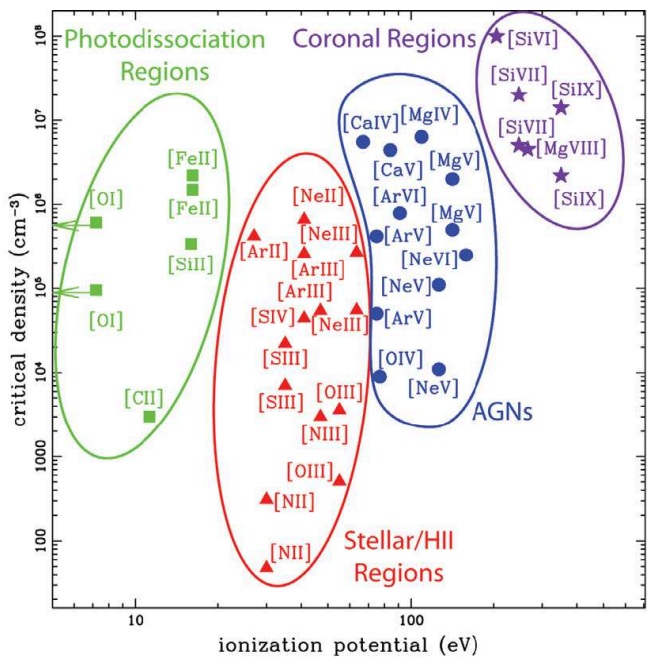}
  \end{center}
  \caption{The positions of the infrared fine structure lines in the density-ionization diagram, showing
the diagnostic potential of these lines to trace different astrophysical conditions: from photodissociation
regions, to Stellar/HII regions, to AGN environments and high excitation coronal line regions \citep{sm92}. }  
\label{spinogliol_fig:fig1}
\end{figure}

\section{MIR/FIR imaging spectroscopy is the key}

Hot and young stars and black hole accretion disks show strong differences in the shape of their ionizing 
continuum. However the far-UV continuum, dominating the total bolometric output luminosity in both processes,  
is in general not directly observable, due to absorption by HI. 
The best tracers and indeed discriminators of accretion and star formation processes are therefore 
emission lines from the photo-ionized gas. In both stabursts and AGN a fairly constant 
fraction (10--20\%) of the ionising continuum gets absorbed by gas and then re-radiated 
as line emission, making this observable a powerful diagnostic tool. 
Detecting the exact fraction of the ionizing radiation absorbed by the gas surrounding the powering 
sources is however not crucial if one uses emission line ratios, which are therefore the most 
efficient observable probes of energy production mechanisms.
To overcome heavy extinction, emission lines in the near-, mid- and far-IR have to be used,
especially to probe obscured regions, such as those dominated by star formation and AGN activity.
{\it MIR/FIR imaging spectroscopy is therefore essential 
to trace galaxy evolution throughout cosmic times in an unbiased way and so to minimize dust extinction.}

\begin{figure}[ht]
  \begin{center}
    \includegraphics[width=\columnwidth]{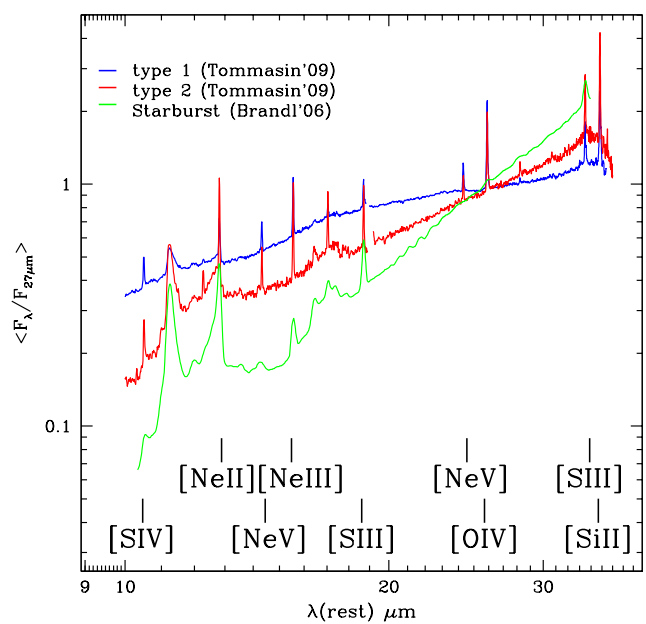}
  \end{center}
  \caption{Mid-IR spectra of Seyfert galaxies in the local Universe \citep{t09}, 
compared to those of starburst galaxies \citep{bra06}.} 
\label{spinogliol_fig:fig2}
\end{figure}

By the launch of SPICA, deep cosmological surveys undertaken by ISO, AKARI and Spitzer 
along with those planned for Herschel and SCUBA-2 
will have produced catalogues containing the fluxes of many tens of thousands of faint MIR/FIR/submm sources. 
Photometric surveys, however, can only be used to determine source counts and
a great amount of multiwavelength photometric data and {\it a priori} assumptions are required for 
obtaining redshifts and luminosities. 
Even more importantly,  {\it photometric data alone do not allow to unambiguously  differentiate between 
AGN and star formation activity.}

Substantial progress in studying galaxy evolution can only be achieved by using spectroscopic surveys.
Having direct MIR/FIR spectroscopy can not only provide us with measured (rather than estimated) 
redshifts, but can also unambiguoulsly characterise the detected sources, measuring the AGN and starburst
contributions to their bolometric luminosities over a wide range of cosmological epochs. 
{\it And this will all happen in a single observation}.  
The {\it great}  difference between the surveys that {\it Herschel} will perform 
and those made possible by an instrument like {\it SAFARI} is precisely this: 
the capability of obtaining {\it unbiased spectroscopic data} in a rest-frame spectral range that has been shown 
in the local Universe: 1) to be unaffected by extinction and 
2) to contain strong unambiguous signatures of both AGN emission and star formation.

Figure~\ref{spinogliol_fig:fig1} shows how the IR fine structure lines cover the two main
physical quantities - density and ionization - characterising photoionized and photon dissociated gas. 
Both star formation and stellar evolution processes as well as black-hole accretion can
be traced from a devoted study of a combination of lines and line ratios. The long wavelength of these lines
ensures that dust extinction is minimized. \citet{sm92} first predicted line intensities of AGN and
starburst galaxies using photoioization codes.
To have an idea of the galaxies  that could be 
observed by SAFARI in the far-IR in the redshift range 0.4$<$z$<$4.0, one has to look at the rich rest frame mid-IR spectra
of local galaxies that have been recently observed by the {\it Spitzer} satellite.
Figure~\ref{spinogliol_fig:fig2} shows the average {\it Spitzer} IRS high-resolution mid-IR spectra \citep{t09} of the 12$\mu$m 
sample of Seyfert galaxies \citep{rms}, compared to those of starburst galaxies \citep{bra06}.
The quality of the data is very high and shows the many features that can differentiate
between AGN and star formation processes.

\section{Are local templates observable at high z ?}
\label{}

The main question that we need to answer, before moving to more detailed modeling,
is if the local galaxies, for which we have detailed knowledge of their IR spectra, can be
observed at high redshifts.
In order to do so, we predicted the line intensities as a function of redshift (in the range z=0.1-5) 
for three local template objects: NGC1068 \citep{ale,s05}, 
a prototypical Seyfert 2 galaxy containing both an AGN and a starburst; 
NGC6240 \citep{lu03}, a bright starburst with an obscured AGN and  
M82 \citep{fo01,co99}, the prototypical starburst galaxy. 
We assumed that the line luminosities scale 
as the bolometric luminosity and adopted a luminosity evolution proportional to (z+1)$^2$, 
consistent with recent {\it Spitzer} results at least up to redshift z=2 \citep{per05}.

\begin{figure}[!h]
  \includegraphics[width=\columnwidth]{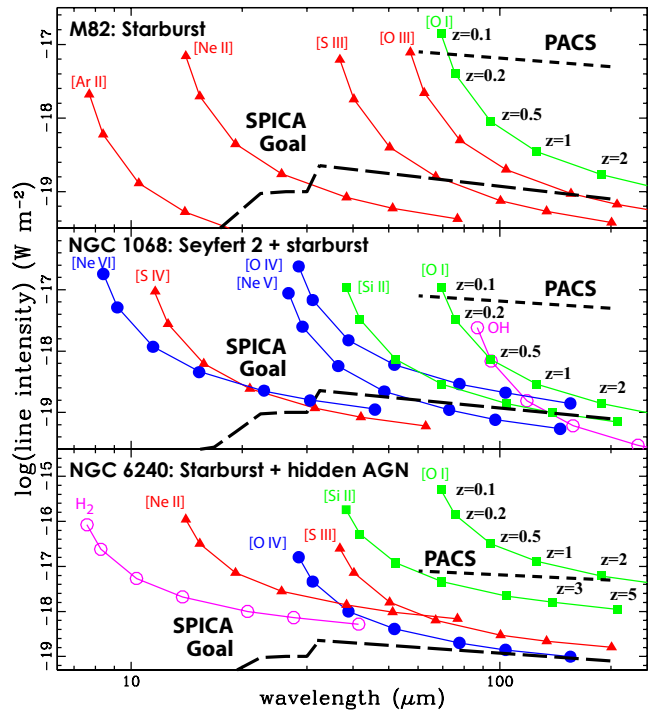}
  \caption{Line observability with PACS onboard of \textit{Herschel} and  with \textit{SPICA}.  Few selected 
diagnostic lines are shown as a function of redshift for the three template objects M82, NGC1068 and NGC6240   
(from top to bottom). The lines have the same symbols as in Figure~\ref{spinogliol_fig:fig1}, 
except for the open circles, which represent molecular lines. Line intensities are given in W m$^{-2}$. 
The short dashed and long-dashed lines respectively give the  5$\sigma$,  1 hour sensitivities of the PACS and SPICA spectrometers.}
  \label{spinogliol_fig:fig3}
\end{figure}

For simplicity\footnote{ We note that the dependence on different cosmological models is not very strong.
The popular model with $\Omega_{M}$= 0.27, $\Omega_{\Lambda}$=0.73, H$_{0}$=71 km s$^{-1}$ Mpc$^{-1}$ would
imply greater dilutions on line intensities, dilutions which
increase with z by a factor of 1.5 for z=0.5 to 2.5 for z=5.}, 
we adopted an Einstein-De Sitter model Universe, with
$\Omega_{\Lambda}$ = 0 and $\Omega_{M}$= 1,
H$_{0}$=75 km s$^{-1}$ Mpc$^{-1}$. The luminosity distances have
been derived using:
$$d_{L} (z)= (2c/H_{0})\cdot [1+z - (1+z)^{1/2}]$$.
The results for the three template objects are reported in 
Figure~\ref{spinogliol_fig:fig3}. 
Among the brightest lines we show the 
[SIV]10.5$\mu$m, the [NeII]12.8$\mu$m and the [OIII]52$\mu$m diagnostic
for the stellar/HII regions,  the [NeV]24.3$\mu$m and the [OIV]25.9$\mu$m 
for the AGN component, the [OI]63$\mu$m and the [SiII]33.5$\mu$m, 
for the photodissociation regions and the OH and H$_2$ rotational lines for the
warm molecular component.

It is clear from the figure that the PACS spectrometer will be able to observe only the most 
favorable object (NGC6240) up to z=2 in the brightest line ([OI]63$\mu$m), while the SPICA 
spectrometer \textit{goal} sensitivities will allow deep infrared spectroscopic studies 
of these relatively low luminosity objects (L$_{IR}$=4, 20, 50 $\times$10$^{10}L_{\odot}$ for
M82, NGC1068 and NGC6240, respectively)
up to z $\sim$ 1-2 for most lines and pushing this limit to even higher redshifts for the brightest lines.

\section{Starburst and AGN at  z=0-4 with SAFARI}
\label{}

With the goal of estimating the number of galaxies at different redshifts that can be detected 
by SAFARI spectroscopic surveys, we have followed the method of: (1) identifying star formation and AGN activity by 
rest frame mid-IR spectroscopic tracers; (2) determining the correlations between line and continuum IR 
luminosity in the Local Universe; (3) using reliable IR continuum luminosity 
functions (derived from both observations and models) and (4) transforming the continuum luminosity functions 
(z=0-4) into line luminosity functions, from which to derive the number of detectable sources
per redshift range.
The use of different models (listed below) allows to provide a reliable range for SPICA-SAFARI predictions.

\begin{figure}[ht]
  \begin{center}
  \includegraphics[width=\columnwidth]{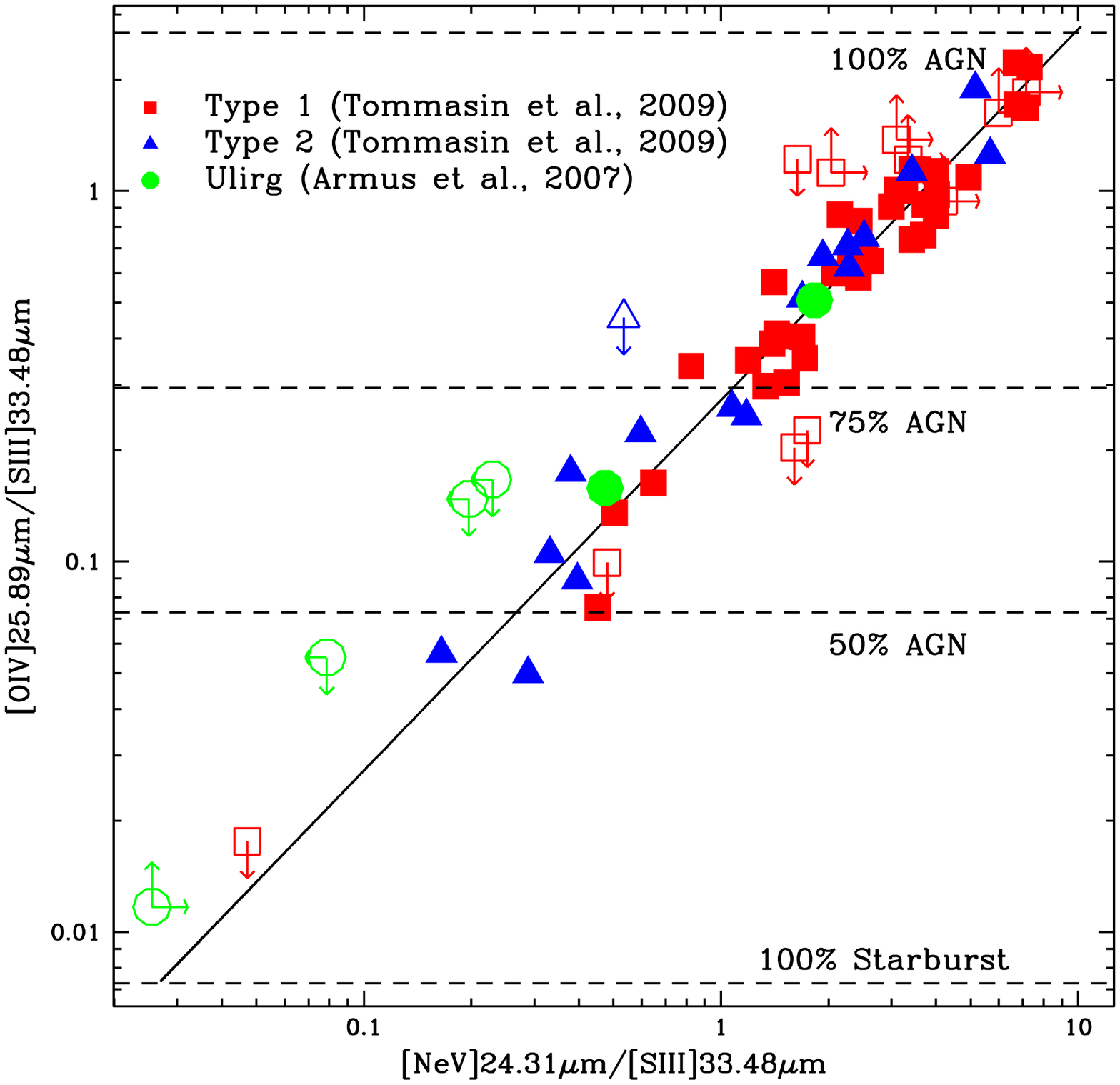}
  \end{center}
  \caption{The [OIV]26$\mu$m/[SIII]33.5$\mu$m {\it vs} [NeV]24.3$\mu$m/[SIII]33.5$\mu$m diagram for
local Seyfert galaxies \citep{t09} and ULIRGs \citep{arm07}.}  
\label{spinogliol_fig:fig4}
\end{figure}

\subsection{Line ratio diagrams for AGN and starbursts}

A suitable line ratio diagram able to separate the emission from the AGN from that of the
starburst can be built by using lines which can be produced {\it only} in the high ionization
environment due to black-hole accretion (such as the [NeV] lines and, at a less extent, [OIV]26$\mu$m),
together with lines generated by processes due to stellar formation and evolution (such
as those which are bright in HII regions: [NeII], [SIII], etc.). However, because we aim at covering
a wide redshift range as possible with the SAFARI spectrometer, we must choose lines which reside at 
long enough wavelengths. So we have concentrated on 
the [OIV]26$\mu$m/[SIII]33.5$\mu$m {\it vs} [NeV]24.3$\mu$m/[SIII]33.5$\mu$m line diagnostic diagram, that we
show in Figure~\ref{spinogliol_fig:fig4}, because all these can be detected in the SAFARI spectral 
range for any redshift between z=0.4 and 5. In this figure we show that such line ratios
can well separate AGN from star formation dominated galaxies, where we have plotted 
the line ratios of the 12$\mu$m selected
Seyfert galaxies \citep{t09} and the ULIRG data \citep{arm07}. 

\subsection{Correlations between line and continuum}

\begin{figure}[!h]
  \includegraphics[width=9.0cm]{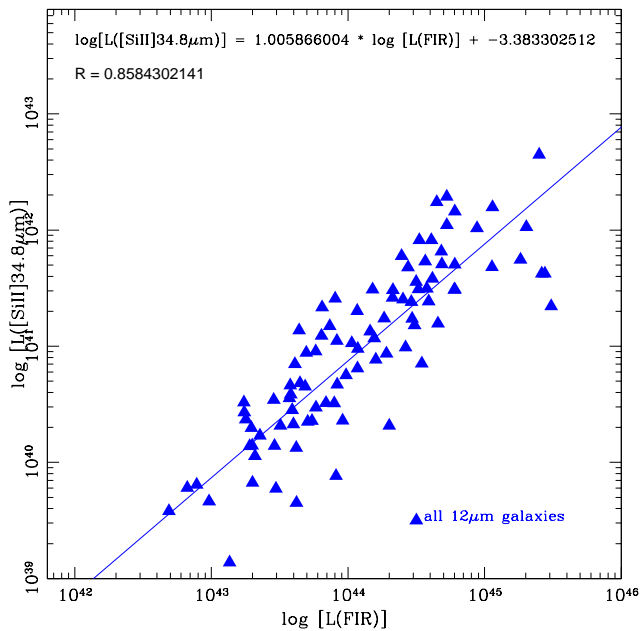}
  \caption{Correlation between the [SiII]34.8$\mu$m and the far-IR luminosity for the complete sample
of 12$\mu$m selected Seyfert galaxies in the local Universe \citep{t08, t09}. }
  \label{spinogliol_fig:fig5}
\end{figure}

\begin{figure}[!h]
  \includegraphics[width=9.0cm]{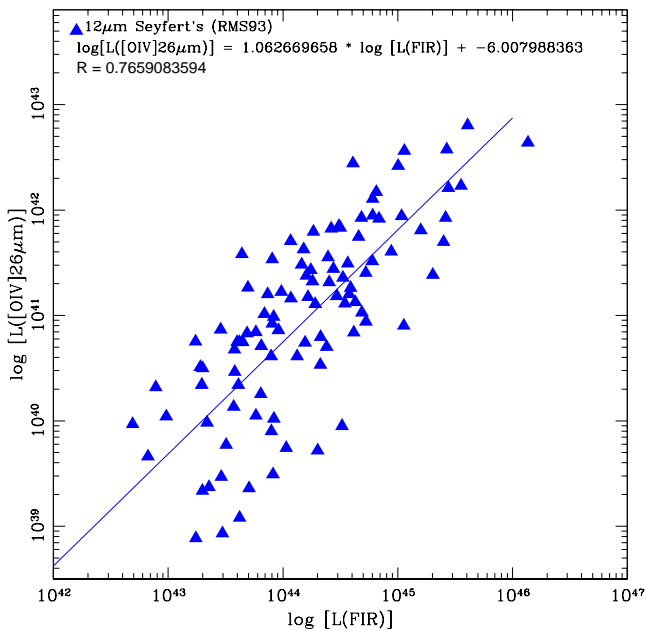}
  \caption{Correlation between the [OIV]26$\mu$m and the far-IR luminosity for the complete sample
of 12$\mu$m selected Seyfert galaxies in the local Universe \citep{t08, t09}.}
  \label{spinogliol_fig:fig6}
\end{figure}

To be able to convert the IR and far-IR continuum luminosity functions into line luminosity functions, 
we have derived the correlations between line and continuum luminosities using the relatively large
and complete sample of local Seyfert galaxies selected at 12$\mu$m which counts for more than 100 objects.
We present in Figure~\ref{spinogliol_fig:fig5} and Figure~\ref{spinogliol_fig:fig6} the 
relation obtainedfor this sample between the far-IR luminosity vs the [SiII]34.8$\mu$m 
and [OIV]26$\mu$m line luminosities, respectively. From such a relation we then derived an 
analytical formula (obtained by least squares fits) to be used in the following analysis.

\subsection{Simulations with the Gruppioni model}

\begin{figure}[!h]
  \includegraphics[width=9.0cm]{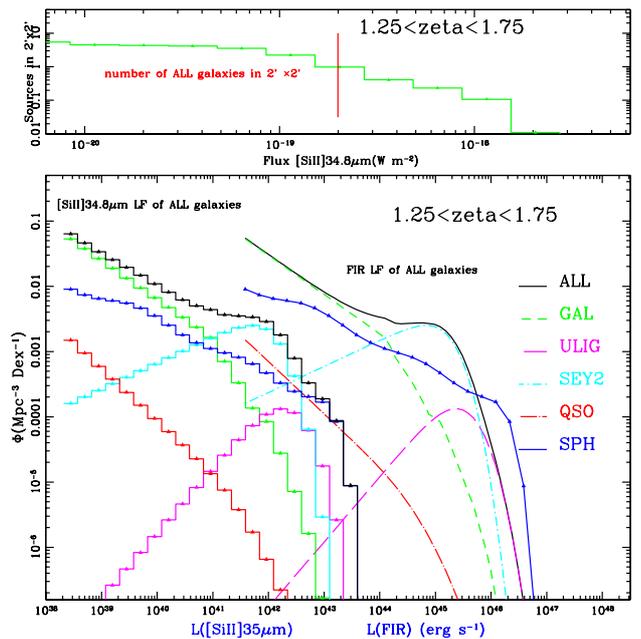}
  \caption{ {\bf Bottom panel:} Composite plot with the FIR luminosity function from \citet{Gru09a}
(right hand side) 
and the derived [SiII]34.8$\mu$m line luminosity function (left hand side) for the various galaxy populations
in the redshift range 1.25$<$z$<$1.75: star forming galaxies, ULIG, Seyfert 2's, Quasars and Spheroids.
{\bf Top panel:} derived total number of sources in a SAFARI field (2' $\times$ 2')
at 1.25$<$z$<$1.75 as a function of the sensitivity. The vertical line shows the 5$\sigma$, 1 hour 
sensitivity limit of 2.0 $\times$ 10$^{-19}$ W m$^{-2}$.}
  \label{spinogliol_fig:fig7}
\end{figure}

Our first method is based on the galaxy evolution model developed 
by Gruppioni and collaborators \citep{Gru09a, Gru09b}, which makes use of all the
available IR data to extrapolate continuum luminosity functions from $z=0$ to $z\sim4$, 
with distinct contributions from starbursts and AGN. To account for the population of spheroids 
(proto-ellipticals) which locally do not appear as substantial far-IR emitting sources, but are expected to 
be relevant at high redshifts (e.g. SMG galaxies), we integrated the model by \citet{Gra04} 
which uses an ab-initio approach for galaxy formation and evolution (see also \citet{lap06})
capable to predict multi-wavelength luminosity functions and number counts. 
As an example, we show in Figure~\ref{spinogliol_fig:fig7} the continuum and line luminosity functions and 
the predicted number of sources detectable in the [SiII]34.8$\mu$m line 
in the redshift range 1.25$<$z$<$1.75, as derived from the modeled FIR luminosity function.

By following this approach, we expect to detect about 7 sources per field of view, over the redshift range 0.$<$z$<$4.0, 
in the [SiII]34.8$\mu$m line, which is a star formation indicator.
We assumed a 5$\sigma$, 1 hour detection limit of 2$\times$ 10$^{-19}~W~m^{-2}$. 
The lines of [OIV]26$\mu$m (indicator of AGN activity) and [SIII]33$\mu$m are, 
in the local galaxy population, a factor 1.5-2 fainter than the [SiII] line, though 
they are found to be of the same order of magnitude in LIRGs and ULIRGs \citep{Vei09}. 
Given the above, we expect to detect all three lines simultaneously in all the observed sources.

\subsection{Simulations with the Franceschini model}

As a second approach we adopted the model developed by Franceschini and collaborators 
at Padova University \citep{fra09}. This is a backward 
evolution model which fits all available data from Spitzer, ISO, COBE, SCUBA, etc. It 
includes direct determinations of multi-wavelength redshift-dependent luminosity 
functions from Spitzer and accounts in great detail not only for star forming 
galaxies, but also for type-1 and type-2 AGNs.

\begin{figure}[!h]
  \includegraphics[width=\columnwidth]{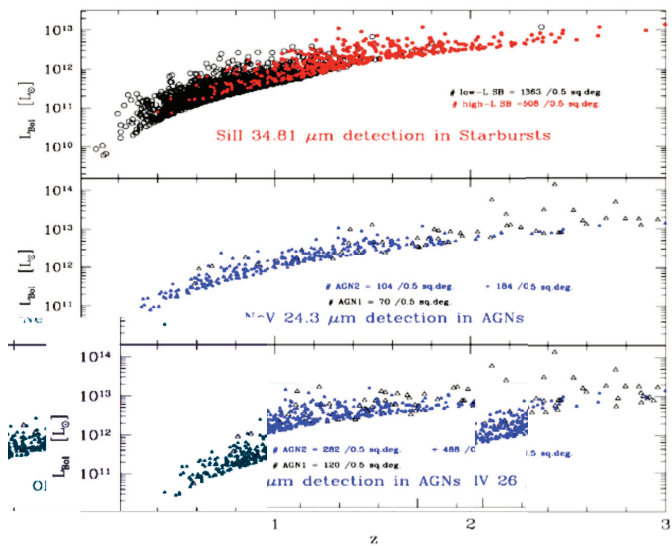}
  \caption{Prediction of the results of a spectroscopic survey over 0.5 square degrees 
(requiring of order of 500 hrs integration with nominal sensitivity) from the model by 
\citet{fra09}.
The number of detected sources would be 120 Type-1 AGNs, 770 Type-2 AGNs and 
1870 starburst galaxies in SiII, for a total of 2800 objects. It has been adopted
the goal sensitivity of SAFARI of 2.0 $\times$ 10$^{-19}$ W m$^{-2}$ (5$\sigma$, 1 hour).}
  \label{spinogliol_fig:fig8}
\end{figure}

The simulation has considered the contributions of: moderate-luminosity starburst galaxies
 and high-luminosity starbursts, for which we have considered the [SiII]34.81$\mu$m line;
Type-1 Active Galactic Nuclei and quasars, for which we have considered the [NeV]24.3$\mu$m 
and [OIV]26$\mu$m lines. Type-2 AGNs also provide substantial contributions to the detection 
statistics (they are considered to be a fraction of the low-L and high-L  SBs)
We referred to the 2.0 $\times$ 10$^{-19}$ W m$^{-2}$  5$\sigma$, 1hr limit.
We predict $\sim$4 starburst, 2 type-2 and 1 type-1 AGN per SAFARI field of view of 2\arcmin $\times$ 2\',
for a total of 7 sources, in good agreement with the predictions obtained in Section 4.3
with a totally independent model.
We show, as an example, in Figure~\ref{spinogliol_fig:fig8} the results of a spectroscopic survey over 0.5 square degrees.

\section{Conclusions}

The main conclusions of this work are:
\begin{itemize}
\item[-] After decades of efforts we are close to having reliable IR measures of  
{\it star formation rate} and  {\it AGN accretion power}, through {\it IR/FIR spectroscopic
surveys}, completely unaffected by dust, allowing us to study the evolution of galaxies in 
terms of the power produced by the main energy production mechanisms.

\item[-] Accurately measuring the power due to stellar formation \& evolution and that due to 
gravity in collapsed nuclei is the first step towards understanding them, 
and how they have been related over the history of the Universe.

\item[-] We learned how to measure these in local galaxies through mid-IR spectroscopy.

\item[-] Blind FIR spectroscopic surveys with SAFARI can be the mean to physically measure galaxy evolution,
provided that a sensitivity of 2$\times$ 10$^{-19}$ W/m$^2$ (5$\sigma$, 1 hour) 
can be reached. 

\item[-] Two different and independent galaxy evolution models predict about 
7-10 sources to be spectroscopically detected in more than one line in a 2'$\times$ 2'
SAFARI field of view, down to the expected flux 
limits of SPICA, with about 20\% of sources to be detected at z$>$2.
SPICA-SAFARI will be therefore excellent at detecting high-z sources and 
at assessing in a direct way their nature (e.g whether mainly AGN or Star 
Formation powered) thanks to blind spectroscopy.
\end{itemize}

\begin{acknowledgements}
This work has been funded in Italy by the Italian Space Agency (ASI) with contract ASI-I/057/08/0.
\end{acknowledgements}

\end{document}